\documentclass{article}

\usepackage{algorithm}
\usepackage{amsmath,subfigure}
\usepackage{arxiv}
\usepackage{algpseudocode}
\usepackage[utf8]{inputenc} % allow utf-8 input
\usepackage[T1]{fontenc}    % use 8-bit T1 fonts
\usepackage{hyperref}       % hyperlinks
\usepackage{url}            % simple URL typesetting
\usepackage{booktabs}       % professional-quality tables
\usepackage{amsfonts}       % blackboard math symbols
\usepackage{nicefrac}       % compact symbols for 1/2, etc.
\usepackage{microtype}      % microtypography
\usepackage{lipsum}
\usepackage{graphicx}
\graphicspath{ {./images/} }

\title{Embedding Is (Almost) All You Need: Retrieval-Augmented Inference for
Generalizable Genomic Prediction Tasks}

\author{
Nirjhor Datta$^{1,2}$ \\
   \And
Swakkhar Shatabda$^{2}$ \\
 \And
 M Sohel Rahman$^{1}$\\
  \AND
  $^{1}$Department of Computer Science and Engineering, \\
  Bangladesh University of Engineering and Technology, \\
  West Palashi, Dhaka 1205, Bangladesh \\
  \And
  $^{2}$Department of Computer Science and Engineering, \\
  BRAC University, Dhaka 1205, Bangladesh \\
}

\begin{document}
\maketitle
\begin{abstract}
Large pre-trained DNA language models such as DNABERT-2, Nucleotide Transformer, and HyenaDNA have demonstrated strong performance on a wide range of genomic benchmarks. However, most applications rely on expensive fine-tuning, which benefits significantly from the similar distribution between training and test data. In this work, we investigate whether task-specific fine-tuning is always necessary. We show that simple embedding-based pipelines  extracting fixed representations from these models and feeding them into lightweight classifiers can achieve competitive performance. Moreover, in independent evaluation settings with different data distributions, embedding-based methods outperform fine-tuning in several tasks while reducing inference time by 10×–20×. Our results suggest that embedding extraction is not only a strong baseline but also a more generalizable and efficient alternative to fine-tuning, particularly for deployment in diverse or unseen genomic contexts. Qualitatively, our retrieval‐augmented pipeline combining fixed transformer embeddings (DNABERT‐2, Nucleotide Transformer, HyenaDNA) with lightweight sequence features achieved competitive result across nine genomic tasks. Also, DNABERT-2 and Hyenadna embeddings with simple sequence features closely matches or exceeds fine-tuned transformer performance on two new independent test set, enhancer and promoter classification tasks, without any model retraining. Quantitatively, our embedding-based approaches consistently outperform fine-tuning in terms of carbon efficiency while maintaining competitive accuracy. For enhancer classification, embedding with zCurve achieves an accuracy of 0.68 using HyenaDNA (vs. 0.58 for fine-tuned HyenaDNA) with almost {88\% reduction in inference time} and over {8$\times$ lower carbon emissions} (0.02 kg vs. 0.17 kg CO$_2$). For non-TATA promoter classification, fixed embeddings from DNABERT-2 combined with zCurve, GC content, or AT/GC ratio reach {0.85 accuracy}, compared to 0.89 for fine-tuning, but with a {22$\times$ lower carbon footprint} (0.02 kg vs. 0.44 kg CO$_2$). Similarly, HyenaDNA with embeddings achieves up to {0.84 accuracy} while emitting just {0.01 kg CO$_2$} compared to 0.04 kg for fine-tuning. These results show embedding-based pipelines can offer over {10$\times$ higher carbon efficiency} while achieving similar or better predictive performance. The code is available here:  \url{https://github.com/NIRJHOR-DATTA/EMBEDDING-IS-ALMOST-ALL-YOU-NEED}.
\end{abstract}

% keywords can be removed

\section{Introduction}
Genomic sequence analysis is one of the most promising areas of research in bioinformatics. The ability to understand the functional elements of the genome, such as promoters, enhancers, and splice sites, has vast implications for fields, such as gene regulation, personalized medicine, and disease prediction \cite{shlyueva2014transcriptional}. Traditional computational methods often rely on handcrafted features or pre-defined models, which, while useful, can struggle to capture the complexities inherent in genomic sequences \cite{zhang2024computational}. Recent advances in transformer‐based models, originally developed for natural language processing, have been successfully adapted to nucleotide sequences, yielding state‐of‐the‐art performance on a variety of benchmarks \cite{ji2021dnabert, dalla2025nucleotide, nguyen2023hyenadna}. 

% DNABERT \cite{ji2021dnabert}, inspired by the BERT \cite{devlin2019bert} architecture from natural language processing (NLP), was the first transformer-based model applied to DNA sequences. This model tokenizes genomic sequences into k-mers \cite{robin2001numerical}, enabling it to capture long-range dependencies and contextual information within DNA sequences. The success of DNABERT \cite{ji2021dnabert} was later expanded with DNABERT-2 \cite{zhou2023dnabert}, which incorporated byte-pair encoding (BPE) and multi-species support, improving both efficiency and generalizability. Despite these advances, most of these models require fine-tuning on the task-specific dataset to achieve optimal performance.

% However, fine-tuning these models on large datasets is computationally expensive and time-consuming, often requiring substantial domain expertise and extensive computational resources. This process is a significant bottleneck in applying deep learning models effectively across multiple genomic tasks and datasets.

% Fine-tuning often leads to over-fitting especially when we work with small datasets, which is common in genomics. 

Despite their empirical success, these transformer‐based methods are computationally expensive to fine‐tune and deploy, often requiring hundreds of GPU hours and substantial memory resources. Prior work has highlighted the environmental impact of large‐scale model training, showing that the carbon footprint of training a single transformer can rival that of multiple automobiles over their lifetimes \cite{strubell2020energy, patterson2021carbon}. In genomics, where sequence lengths frequently exceed 10\,Kb, the quadratic complexity of self‐attention exacerbates these challenges, making full fine‐tuning impractical in many research and clinical settings.

To address these limitations, retrieval‐augmented methods, originally proposed in the NLP domain offer a promising alternative. By indexing fixed transformer embeddings in FAISS (Facebook AI Similarity Search) and performing nearest‐neighbor retrieval at inference time, one can sidestep full model fine‐tuning while still leveraging rich contextual representations \cite{lewis2020retrieval, johnson2019billion}. Recent works in gene interaction prediction has shown that retrieval‐augmented generation can achieve better accuracy with substantially lower computational cost \cite{lin2024generag}. In parallel, hybrid approaches that combine pretrained embeddings with handcrafted genomic features (e.g., GC content, z‐curve, pseudoKNC) have demonstrated marginal gains in classification accuracy while maintaining lightweight inference. So, there is a need for an approach which can levarage the power of large pre-trained models, without the need for fine-tuning, but still achieve state of the art results on various genomic classification tasks. This motivates the exploration of alternative methods that can provide high performance without the computational cost and complexities associated therewith. 

In this work, we propose a comprehensive, retrieval‐augmented genome classification framework that: (1) employs pretrained transformer embeddings (DNABERT‐2, Nucleotide Transformer, HyenaDNA) as fixed representations; (2) optionally augments these embeddings with biologically motivated features such as GC content, z‐curve components, AT/GC ratio, cumulative skew, and pseudoKNC; and (3) uses FAISS $L_2$ indexing for fast k‐nearest‐neighbor retrieval and weighted voting. 
We have evaluated our approach on nine diverse genomic benchmarks. The central focus of this research is to investigate whether fixed, pretrained embeddings can serve as a competitive alternative to full model fine-tuning, particularly in scenarios that demand strong generalizability. To this end, we conduct evaluations on a new, independent test set to assess how well the learned representations transfer beyond the training distribution. Our embedding-based approach achieves comparable or even superior performance to fine-tuned models, while reducing GPU fine-tuning time by over 70\% and reducing carbon emission by upto 77.5x, thereby promoting a more sustainable and efficient paradigm aligned with Green AI principles \cite{schwartz2020green}.

% \subsection{Research Contribution}
% In this study, we present a novel approach that combines the power of pre-trained genomic models with a retrieval-augmented mechanism, enabling us to achieve near state-of-the-art performance without the need for fine-tuning. 
The key contributions of this work are as follows.
% \begin{enumerate}
% \item{We demonstrate that the integration of Retrieval-Augmented Generation (RAG) with pre-trained genomic models, such as DNABERT or DNABERT-2, allows for accurate genomic sequence classification without any task-specific fine-tuning. This approach eliminates the need for computationally expensive and time-consuming fine-tuning procedures.}

% \item{The proposed method is highly scalable, as it can be applied to multiple genomic classification tasks without requiring task-specific fine-tuning. Furthermore, it is flexible in its ability to incorporate various genomic sequence models and external databases for a range of classification challenges.}
% \end{enumerate}
\begin{enumerate}
    \item \textbf{Retrieval‐Augmented Classification Pipeline:} We propose a hybrid framework that combines pretrained transformer embeddings (DNABERT‐2, Nucleotide Transformer, HyenaDNA) with handcrafted genomic features (e.g., pseudoKNC, GC‐content, z‐curve) and a lightweight kNN classifier. By indexing precomputed embeddings in FAISS and using majority/weighted voting, we avoid full fine‐tuning of large models at inference time, drastically reducing GPU memory usage and compute cost.

    % \item \textbf{Transformer‐Level Accuracy with Lightweight Methods:} Empirically, our retrieval‐augmented approach achieves near‐state‐of‐the‐art performance on nine diverse genomic benchmarks. For example, fine‐tuned Nucleotide Transformer reaches up to 0.98 AUROC on species discrimination (Human vs Worm) and 0.96 AUROC on coding vs Intergenomic classification, while pure embedding‐only methods (e.g., DNABERT‐2 + pseudoKNC) achieve up to 0.87 AUROC on coding vs Intergenomic, demonstrating that simple feature‐fusion combined with retrieval can closely match heavy fine‐tuning.

    % \item \textbf{Zero‐Shot Adaptability:} Our framework supports zero‐shot classification by retrieving nearest neighbors from an existing index without retraining.

    % \item\textbf{Evaluating performance on independent test set:} We evaluated our approach on an independent test set across two genome classification tasks, using both a fine-tuned model trained on a genomic benchmark dataset and a retrieval-based method that leverages embeddings from the same dataset. We report both classification accuracy and inference time for these two tasks.
    
    % \item \textbf{Long and Ultra‐Long Sequence Support:} Unlike standard transformer architectures limited to input lengths of 512–1024 tokens, our retrieval mechanism handles entire genomic contigs without chunking.

   \item\textbf{Modular, Scalable, and Carbon-Efficient Architecture:} The embedding and retrieval stages are decoupled, enabling plug-and-play substitution of any pretrained model or handcrafted feature set. This modularity allows flexible trade-offs between accuracy and resource usage, with our pipeline achieving up to {77.5$\times$ lower} CO\textsubscript{2} emissions compared to full fine-tuning while maintaining competitive performance.

    \item \textbf{Comprehensive Empirical Evaluation:} We benchmark our method on nine publicly available genomic datasets (e.g., Human Ensembl Regulatory, Drosophila Enhancers Stark, Human Non-TATA Promoters), comparing embedding‐only, feature‐augmented, and fine‐tuned variants across three transformer models. We evaluated our approach on an independent test set across two genome classification tasks, using both a fine-tuned model trained on a genomic benchmark dataset and a retrieval-based method that leverages embeddings from the same dataset.

\end{enumerate}
\section{Related Work}

Transformer architectures have revolutionized sequence modeling across domains, including genomics. DNABERT \cite{ji2021dnabert} introduced a BERT-style model pre-trained on k-mers from the human genome, achieving strong performance on promoter and enhancer classification. However, limitations such as redundant k-mer tokenization, short context windows (512 tokens), and single-species training motivated DNABERT-2 \cite{zhou2023dnabert}, which uses byte pair encoding, multi-species data, and longer input handling via gradient checkpointing.

Nucleotide Transformer (NT) \cite{dalla2025nucleotide} extends this trend by pre-training large-scale transformer models (up to 2.5B parameters) on thousands of genomes. NT embeddings support strong performance across diverse tasks even in low-data settings, demonstrating the utility of context-aware nucleotide representations.

HyenaDNA \cite{nguyen2023hyenadna}, inspired by efficient implicit convolution mechanisms, offers longer context windows with lower time complexity. It outperforms existing methods on multiple GenomicBenchmarks tasks while using fewer parameters and resources.

Despite these advances, fine-tuning large models remains energy-intensive. Studies by Strubell \cite{strubell2020energy} and Patterson \cite{patterson2021carbon} highlight the environmental cost of deep learning, advocating for energy-efficient alternatives. In response, Green AI \cite{schwartz2020green} promotes trade-offs between performance and resource consumption using metrics like FLOPs, GPU hours, and emissions.

Retrieval-Augmented Generation (RAG) \cite{lewis2020retrieval} has emerged as a lightweight alternative, replacing full model updates with external memory lookups via FAISS \cite{johnson2019billion}. GeneRAG \cite{lin2024generag} applied this idea to genomics, boosting performance without costly fine-tuning.

While transformer embeddings have been shown to cluster functionally related sequences \cite{ji2021dnabert, dalla2025nucleotide,nguyen2023hyenadna}, few works compare retrieval-based classification across multiple models and tasks. We address this by benchmarking DNABERT-2, HyenaDNA, and NT under a unified retrieval framework, augmented with biologically meaningful descriptors, offering an efficient and accurate alternative to fine-tuning.

\section{Proposed Method}
\label{methods}
\subsection{Overview}
\begin{figure*}[ht]
    \centering
    \includegraphics[width=1.1\linewidth]{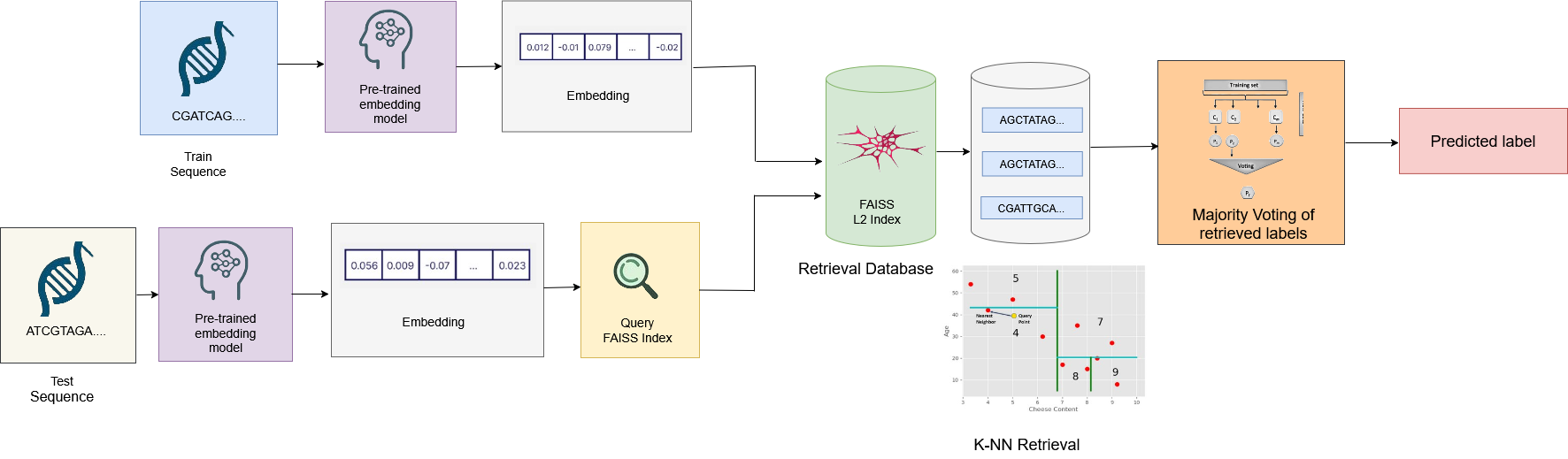}
    \caption{Top-level overview of the proposed framework}
    \label{fig:method_overview}
\end{figure*}

We propose a novel retrieval-augmented genomic classification framework that eliminates the need for fine-tuning large genomic language models. First, raw DNA sequences are embedded using a pretrained model (such as, DNABERT-2 or Nucleotide Transformer) without any parameter updates. During inference, a nearest-neighbor retrieval mechanism is employed to fetch relevant training embeddings based on similarity. The retrieved context is combined with the test sequence representation. Classification is performed directly by comparing similarity scores, avoiding the need for additional fully connected layers or retraining. This lightweight approach enables efficient zero-shot or few-shot classification, particularly suitable for imbalanced datasets. Throughout the rest of the paper, we define this pipeline as {base framework}.
% Our method demonstrates competitive or near state-of-the-art results without extensive computational overhead. 
Top level diagram of our proposed method is presented at Figure \ref{fig:method_overview}. 
Our proposed framework leverages a training-free, retrieval-augmented classification paradigm for genomic sequence classification. The methodology involves three core components: (1) hybrid feature extraction, (2) retrieval using FAISS similarity search, and (3) label prediction via weighted voting. This approach eliminates the need for training complex models while maintaining competitive classification performance.
\subsection{Hybrid Feature Extraction}

Each DNA sequence is first encoded using two complementary representations as follows:

    \textbf{Pretrained Embeddings.} We utilize embeddings obtained from a pretrained transformer model trained on genomic sequences. These embeddings capture the high-level contextual semantics of the sequences.
    \begin{equation}
    \mathbf{e}_i = f_{\theta}(\mathbf{x}_i) = \frac{1}{L_i} \sum_{t=1}^{L_i} h_t,
\end{equation}
where $f_{\theta}$ is the pretrained transformer model with frozen weights and $h_t$ are the hidden representations of each nucleotide position $t$ in the sequence $\mathbf{x}_i$.

     \textbf{Biological Feature Engineering.} Inspired by PyFeat \cite{muhammod2019pyfeat}, to complement the high-dimensional representations from pretrained embeddings, we extract set of biologically-inspired handcrafted features that capture essential patterns and characteristics of DNA sequences, namely, AT/GC Ratio
, Z-curve features
, Cumulative GC and AT skew and
 Pseudo k-mer composition.

The learned embedding vector \( \mathbf{e}_i \) is \(\ell_2\)-normalized and linearly weighted by a coefficient \( \alpha \), while the handcrafted feature vector \( \mathbf{f}_i \) is min-max normalized and weighted by \( \beta \). These scaled representations are concatenated to form the final hybrid feature vector:
\[
\mathbf{h}_i = \alpha \cdot \frac{\mathbf{e}_i}{\|\mathbf{e}_i\|} \mathbin{\|} \beta \cdot \text{MinMax}(\mathbf{f}_i),
\]
where \( \| \) denotes vector concatenation. We found that using \( \alpha = 0.8 \) and \( \beta = 0.2 \) yielded better performance compared to other combinations of weights.

These handcrafted features are concatenated and min-max normalized before fusing with deep embeddings to form the final hybrid representation used for similarity search and classification.

\subsection{Similarity-Based Retrieval}

Given the hybrid representation of the training data, we build a similarity search index using the FAISS library, which supports efficient nearest neighbor search based on inner product similarity. For a test sequence with hybrid vector $\mathbf{h}_q$, the system retrieves its top-$k$ most similar training samples from the index. A dynamic $k$ retrieval strategy is adopted, where $k$ is increased adaptively if the mean similarity of the top-$k$ results falls below a predefined threshold $\tau$.
\begin{equation}
    k = \min \left\{ k' : \frac{1}{k'} \sum_{j=1}^{k'} s_j \ge \tau \right\}.
\end{equation}
\subsection{Weighted Voting for Prediction}

Let $\mathcal{N}_q = \{(\mathbf{h}_{i_j}, y_{i_j}, s_j)\}_{j=1}^k$ denote the retrieved neighbors, where $y_{i_j}$ is the label and $s_j$ the similarity score; a sigmoid transformation is applied to convert similarity scores into soft weight, $w_j = \frac{1}{1 + e^{-s_j}}$.

These weights are then accumulated per class to determine the final prediction:
\[
\hat{y}_q = \arg\max_{c} \sum_{j=1}^k w_j \cdot \mathbb{I}[y_{i_j} = c]
\]

% \subsection{Advantages}

% This methodology offers the following benefits:

% \textbf{Training-free} No backpropagation or weight updates are required.

%  \textbf{Efficient and scalable} FAISS provides fast and scalable approximate nearest neighbor search.
 
%  \textbf{Interpretability} The influence of each neighbor on the final prediction can be easily traced.
\section{Experimental Setup}
\label{}
First, we evaluate our proposed framework along with aditional handcrafted features on genomic benchmark. Even though the GenBench paper \cite{liu2024genbench} presents fine-tuning results on the Genomic Benchmark \cite{grevsova2023genomic}, it does not provide the specific hyperparameters used for those models. As a result, a direct comparison with our approach is not feasible. However, our primary objective is not merely to achieve competitive performance with fine-tuned models, but also to evaluate the generalizability of our method. Therefore, our main experiment focuses on assessing how well our approach performs on a different distribution of data for the same task, highlighting its robustness and effectiveness in generalization and compare the results against fine-tuned based approach in a similar setting. This is why, in the next stage, we evaluate DNABERT-2 and Hyenadna using both our proposed framework and standard fine-tuning on two downstream genomics tasks, namely, {enhancer classification} and {non-TATA promoter classification} using two new independent test sets. Due to its substantially larger parameter count (2.5B) compared to DNABERT-2 (117M) and HyenaDNA (6.5M), the Nucleotide Transformer was excluded from evaluation at this stage. All the models are trained on a A6000 GPU 48 GB,
RAM: 64GB DDR5 SSD. 

\subsection{Datasets}
We evaluate our models on nine publicly available genomic benchmark datasets (Table \ref{tab:dataset_summary}) covering diverse biological classification tasks. These datasets span enhancer detection, promoter classification, and other regulatory sequence identification tasks in various organisms including humans, mice, worms, and Drosophila. All datasets are sourced from the {Genomic Benchmark} \cite{grevsova2023genomic}. 
% A detailed description of the dataset is presented at Table \ref{tab:dataset_summary}. 
% t-SNE map of Genomic DNABERT2 Embeddings across different samples of all datasets is presented at (Supplementary Section Figure 1).
\begin{table}[htbp]
\centering
\caption{Summary of Genomic Datasets Used in Experiments ($|$C$|$ indicates number of classes in the dataset). }
\label{tab:dataset_summary}
\begin{scriptsize}
\begin{tabular}{|c|l|p{1cm}|p{1cm}|l|}
\hline
\textbf{\#} & \textbf{Dataset Name} & \textbf{Train Samples} & \textbf{Test Samples} & \textbf{$|$C$|$} \\
\hline
1 & Human Ensembl Regulatory         & 231,348  & 57,713   & 3 \\
2 & Drosophila Enhancers Stark       & 5,184    & 1,730    & 2 \\
3 & Demo Coding vs Intergenomic Seqs & 75,000   & 25,000   & 2 \\
4 & Demo Human vs Worm               & 75,000   & 25,000   & 2 \\
5 & Human Enhancers Cohn             & 20,843   & 6,948    & 2 \\
6 & Human Enhancers Ensembl          & 123,872  & 30,972   & 2 \\
7 & Human OCR Ensembl                & 139,804  & 34,952   & 2 \\
8 & Human Non-TATA Ensembl           & 27,097   & 9,034    & 2 \\
9 & Mouse Enhancers Ensembl          & 968      & 242      & 2 \\
\hline
\end{tabular}
\end{scriptsize}

\end{table}

For our second and main task, {checking the performance of embedding based mechanisms agains fine-tuned based models}, we have chosen two downstream tasks, namely, human enhancer classification and {human non-tata promoter classification}. We have used {human enhancer ensemble} and {human non-tata promoters} from genomic benchmark as train set. For {human enhancer classification} task, we have evaluated ienhancer \cite{nguyen2019ienhancer} test set and {human non-tata promoter classification} task, dataset described in \cite{umarov2017recognition}  as an independent test dataset. For the \textit{ienhancer} test set, we have merged strong enhancers and week enhancers as enhancer keeping non-enhancer as it is since our original \textit{human enhancer ensemble} dataset from genomic benchmark has only 2 labels - enhancer and non-enhancer.

\subsection{Preprocessing for Fine-Tuning-Based Models}
We convert DNA sequences into k-mer representations (k=6) to capture local nucleotide patterns. These k-mers are then tokenized using the corresponding model tokenizer with a maximum input length of 512. Labels are mapped and datasets are formatted into PyTorch tensors for compatibility with the HuggingFace Trainer API.
\subsection{Model Variants}

We evaluate the following model configurations:

\begin{itemize}
    \item \textbf{Embedding (\textbf{emb})}: Sequence is encoded using pre-trained bilogical language models embeddings without fine-tuning.
    \item \textbf{+ Feature}: Inspired by PyFeat \cite{muhammod2019pyfeat}, to complement the high-dimensional representations from pretrained embeddings, we extract set of biologically-inspired handcrafted features that capture essential patterns and characteristics of DNA sequences. The detailed way of calculating additional features are available at
    \item \textbf{DNABERT-2 (Fine-tuned)}: Full model fine-tuned end-to-end on the training set. Fine-tuning parameters on new independent datasets are presented at Table \ref{tab:hyperparameters}.
% Table 1: Hyperparameter Settings
\begin{table}[ht]
\centering
\caption{Hyperparameter Settings}
\label{tab:hyperparameters}
\begin{tabular}{|l|l|}
\hline
\textbf{Hyperparameter} & \textbf{Value} \\
\hline
Learning Rate           & $2 \times 10^{-5}$ \\
Batch Size (per device) & 32 \\
Number of Epochs        & 3 \\
Weight Decay            & 0.01 \\
\hline
\end{tabular}
\end{table}
\end{itemize}

All hybrid models use a simple KNN  classifier on top of the combined feature vector.
\subsection{Optimized Retrieval Pipeline}

To accelerate the inference stage of our retrieval-based evaluation, we implemented an optimized version of our pipeline with the following key improvements:

\begin{itemize}
    \item \textbf{Batch Processing:} Instead of computing embeddings one at a time, we use a PyTorch DataLoader to process input sequences in batches, significantly improving GPU utilization and memory efficiency.

    \item \textbf{Mean Pooling:} Token-level outputs from the HyenaDNA model are pooled using mean pooling to obtain fixed-length embeddings, replacing any earlier simplistic representations.

    \item \textbf{Efficient Retrieval:} The brute-force distance computation is replaced by a FAISS IndexFlatL2 index, enabling much faster nearest neighbor search on the CPU.

    \item \textbf{Inference Time Tracking:} A timing mechanism is added to measure the total runtime of the retrieval pipeline, providing a quantitative measure of efficiency gains.
\end{itemize}

All other steps, including the use of majority voting over the top-$k$ retrieved labels, remain unchanged from the original implementation. This optimized pipeline was used in the second and main set of experiments, which evaluated the performance of embedding-based methods against fine-tuned models. These optimizations significantly reduced retrieval latency while preserving prediction quality.

% \textbf{DNABERT Limitation:} While we applied this optimized pipeline successfully for HyenaDNA, we were unable to do so for DNABERT due to compatibility issues. The updated version of DNABERT relies on the \href{https://github.com/openai/triton}{Triton} compiler, which is currently unsupported on Windows systems. Therefore, we report DNABERT results from the earlier (unoptimized) pipeline. This affects runtime but not the retrieval quality or accuracy.

% \subsection{Code, Environment and Availability}
% \label{code}
% % All the codes and datasets related to this research are available here. 
% All
% the models are trained on a A6000 GPU 48 GB, RAM: 64GB DDR5 SSD: 1
% TB PSU: 1000W PC. The code is available here:  \url{https://github.com/NIRJHOR-DATTA/RAG-to-Riches-A-Retrieval-Based-Alternative-to-Transformer-Models-in-Genomics}. 
\section{Results}

\subsection{Performance on Genomic Benchmark}
We have presented the algorithms for our base framework (Figure  \ref{fig:method_overview}) and the framework along with handcrafted features.
Top 1\% accuracy of base  and fine-tuned models of DNABERT-2 (117 M), Neucliotide Transformer (2.5 B) and HyenaDNA (6.5 M) across all the datasets of genomic benchmark is presented at Table \ref{tab:full_results}. 

\begin{itemize}
\item\textbf{DNABERT‐2 Embeddings} Accuracy ranges from 0.58 (Human Ensembl Regulatory) to 0.91 (Demo Human vs Worm). Notably, DNABERT‐2 embeddings achieve 0.85 on Demo Coding vs Intergenomic and 0.88 on Human Non‐TATA Ensembl.

\item\textbf{Nucleotide Transformer Embeddings} Accuracy ranges from 0.57 (Human Ensembl Regulatory) to 0.90 (Demo Human vs Worm). In many cases (e.g., Demo Coding vs Intergenomic at 0.85, Demo Human vs Worm at 0.90), its performance closely matches DNABERT‐2.

\item\textbf{HyenaDNA Embeddings} Accuracy spans 0.61 (Drosophila Enhancers Stark) to 0.83 (Demo Coding vs Intergenomic and Demo Human vs Worm). Across datasets, HyenaDNA embeddings lag behind DNABERT‐2 and Nucleotide Transformer by 0.1 - 0.2 AUROC points on average.
\end{itemize}

These results indicate that pretrained transformer embeddings particularly those specialized for genomic sequences encode discriminative features even without task‐specific training. HyenaDNA’s relatively lower embedding performance suggests that its architectural modifications (e.g., convolutional attention mechanisms for long contexts) may not align as directly with the classification tasks in GenBench. The results presented in Table \ref{tab:full_results} are comparable to those reported in the literature with heavy fine-tuning; however, due to the lack of detailed fine-tuning configurations in the GenBench \cite{liu2024genbench}, a direct comparison could not be performed.

\begin{table*}[!htb]
    \centering
    \caption{Top 1\% Accuracy Performance of Embedding Variants and Fine‐Tuned Models Across Nine Genomic Datasets. \textbf{Emb} means the embedding of that particular model. Best scores are in \textbf{bold} }
    \label{tab:full_results}
    \begin{tabular}{lccccccc}
    \toprule
    \multicolumn{8}{c}{\textbf{DNABERT-2}} \\
    \midrule
    \textbf{Dataset}                  & \textbf{Emb} & \textbf{+gcCont} & \textbf{+zCurve} & \textbf{+AT/GC} & \textbf{+cumSkew} & \textbf{+pseudoKNC}  \\
    \midrule
    Human Ensembl Regulatory          & 0.58 & 0.59 & 0.57 & 0.59 & 0.59 & \textbf{0.60}  \\
    Drosophila Enhancers Stark        & 0.70 & \textbf{0.71} & 0.68 & 0.70 & 0.68 & 0.68   \\
    Demo Coding vs Intergenomic         & 0.85 & 0.84 & 0.86 & 0.86 & 0.84 & \textbf{0.87}  \\
    Demo Human vs Worm                & 0.91 & \textbf{0.92} &\textbf{0.92} & \textbf{0.92} & 0.91 & \textbf{0.92}  \\
    Human Enhancers Cohn              & 0.70 & 0.71 & 0.72 & 0.72 & 0.72 & 0.72  \\
    Human Enhancers Ensembl           & 0.71 & 0.72 & \textbf{0.73} & \textbf{0.73} & \textbf{0.73} & \textbf{0.73}  \\
    Human OCR Ensembl                 & 0.63 & 0.64 & \textbf{0.65} & 0.64 & 0.64 & \textbf{0.65}  \\
    Human Non-TATA Ensembl            & 0.88 & 0.88 & 0.86 & 0.87 & 0.88 & 0.83  \\
    Mouse Enhancers Ensembl           & 0.71 & 0.71 & 0.79 & 0.74 & 0.73 & \textbf{0.75} \\
    \midrule
    \multicolumn{8}{c}{\textbf{Nucleotide Transformer}} \\
    \midrule
    \textbf{Dataset}                  & \textbf{Emb} & \textbf{+gcCont} & \textbf{+zCurve} & \textbf{+AT/GC} & \textbf{+cumSkew} & \textbf{+pseudoKNC} \\
    \midrule
    Human Ensembl Regulatory          & 0.57 & 0.57 & 0.55 & 0.57 & 0.57 & \textbf{0.58} \\
    Drosophila Enhancers Stark        & 0.66 & \textbf{0.67} & \textbf{0.67} & 0.66 & 0.66 & 0.65  \\
    Demo Coding vs Intergenomic         & 0.85 & \textbf{0.86} & 0.85 & 0.85 & 0.85 & \textbf{0.86}  \\
    Demo Human vs Worm                & \textbf{0.90} & 0.88 & 0.87 & 0.88 & 0.87 & 0.89 \\
    Human Enhancers Cohn              & 0.66 & \textbf{0.68} & \textbf{0.68} & \textbf{0.68} & \textbf{0.68} & \textbf{0.68} \\
    Human Enhancers Ensembl           & 0.68 & \textbf{0.69} & \textbf{0.69} & \textbf{0.69} & \textbf{0.69} & \textbf{0.69} \\
    Human OCR Ensembl                 & 0.60 & 0.61 & 0.61 & 0.61 & 0.61& \textbf{0.62} \\
    Human Non-TATA Ensembl            & 0.80 & \textbf{0.81} & 0.80 & 0.80 & \textbf{0.81} & \textbf{0.81}  \\
    Mouse Enhancers Ensembl           & 0.78 & 0.75 & \textbf{0.79} & 0.76 & \textbf{0.79} & 0.76 &  \\
    \midrule
    \multicolumn{8}{c}{\textbf{HyenaDNA}} \\
    \midrule
    \textbf{Dataset}                  & \textbf{Emb} & \textbf{+gcCont} & \textbf{+zCurve} & \textbf{+AT/GC} & \textbf{+cumSkew} & \textbf{+pseudoKNC}   \\
    \midrule
    Human Ensembl Regulatory          & \textbf{0.78} & 0.74 & 0.69 & \textbf{0.78} & 0.66 & 0.69 \\
    Drosophila Enhancers Stark        & 0.61 & 0.59 & 0.53 & \textbf{0.63} & 0.52 & 0.51    \\
    Demo Coding vs Intergenomic         & \textbf{0.83} & 0.78 & 0.78 & \textbf{0.83} & 0.78 & \textbf{0.83 }   \\
    Demo Human vs Worm                & \textbf{0.83} & 0.70 & 0.64 & \textbf{0.83} & 0.71 & 0.77  \\
    Human Enhancers Cohn              & 0.67 & 0.65 & \textbf{0.68} & \textbf{0.68} & \textbf{0.68} & 0.67   \\
    Human Enhancers Ensembl           & 0.70 & 0.67 & 0.64 & 0.72 & 0.63 & \textbf{0.73}   \\
    Human OCR Ensembl                 & 0.62 & 0.59 & 0.60 & 0.63 & 0.57 & \textbf{0.63 }   \\
    Human Non-TATA Ensembl            & \textbf{0.82} & 0.71 & 0.73 & 0.79 & 0.73 & 0.74   \\
    Mouse Enhancers Ensembl           & 0.73 & 0.76 & 0.73 & 0.73 & \textbf{0.77} & 0.73   \\
    \bottomrule
    \end{tabular}
\end{table*}
\subsection{Comparative performance on downstream tasks}
We have chosen two down stream tasks from genomic benchmark, namely, {human enhancer classification} and {human non-tata promoter classification}. 
% We have evaluated the performance of both fine-tuned based method and our framework which has used the training set on a new independent dataset.
We evaluated both the fine-tuned model and our proposed framework, trained on the same training set {(human enhancer en-
semble and human non-tata promoters from Genomic Benchmark)}, using a new independent test dataset to assess their generalization performance.We have used DNABERT-2 and Hyenadna with both fine-tuning and our RAG based approach (Suppl. Section Algorithm in the Appendix). 

% For \textbf{human enhancer classification} task, we have evaluated ienhancer \cite{nguyen2019ienhancer} test set and \textbf{human non-tata promoter classification} task, dataset described in \cite{umarov2017recognition}  as an independent dataset. For the \textbf{ienhancer} test set, we have merged strong enhancers and week enhancers as enhancer keeping non-enhancer as it is since our original \textbf{human enhancer ensemble} dataset from genomic benchmark has only 2 labels - enhancer and non-enhancer. We have used DNABERT-2 with both fine-tuning and our RAG based approach (Suppl. Section \ref{algo} in the Appendix). 

For the embedding-based approach, we first extracted sequence embeddings using a frozen DNABERT-2 model and combined them with various handcrafted features, including GC content, zCurve components, AT/GC ratios, cumulative nucleotide skews, and pseudo k-tuple nucleotide compositions (PseudoKNC). Each variant represents a hybrid model that integrates DNABERT-2 embeddings with one specific feature type. In contrast, the fine-tuned baseline directly updates the DNABERT-2 model weights on the downstream task through supervised fine-tuning.

We report both accuracy and inference time (including feature extraction) on a held-out independent test set to assess generalization and computational efficiency. Results on additional independent datasets are provided in Table~\ref{tab:merged_classification_results}, where {accuracy} and {carbon emissions} are denoted as {Acc} and {CE}, respectively. For fine-tuned models, the reported time includes preprocessing, fine-tuning, and inference.

\begin{table}[!htb]
\centering
\scriptsize
\caption{Performance and Inference Time for Enhancer and Non‐TATA Promoter Classification Using DNABERT‐2 and HyenaDNA Based Methods}
\label{tab:merged_classification_results}
\begin{tabular}{|p{0.75cm}|p{0.8cm}|l|p{0.5cm}|p{1cm}|p{0.3cm}|}
\hline
\textbf{Task} & \textbf{Model Type} & \textbf{Model} & \textbf{Acc} & \textbf{Inference Time (s)} & \textbf{CE (kg)} \\
\hline
{Enhancer} 
  & {DNA} 
    & Embedding & 0.65 & \textbf{525.74} & 0.02\\
  &BERT-2 & + GC Content & 0.65 &  + 2.61 & 0.02 \\
  & & + zCurve & 0.65 &  + 2.49 & 0.02\\
  & & + AT/GC Ratio & \textbf{0.67} &  + 2.24 & 0.02\\
  & & + Cumulative Skew & \textbf{0.67} &  + 2.29 & 0.02 \\
  & & + PseudoKNC & \textbf{0.67} &  + 10.97 & 0.02\\
  & & Fine‐tuned & 0.61 & 31497.08 & 1.55 \\
  \cline{2-6}
  & {Hyena} 
    & Embedding & 0.65 & \textbf{410.96} & 0.02\\
  &DNA & + GC Content & 0.61 &  + 2.05 & 0.02 \\
  & & + zCurve & \textbf{0.68} &  + 1.94 & 0.02\\
  & & + AT/GC Ratio & 0.65 &  + 1.78 & 0.02\\
  & & + Cumulative Skew & 0.64 &  + 1.82 & 0.02 \\
  & & + PseudoKNC & 0.65 & +10.11 & 0.02\\
  & & Fine‐tuned & 0.58 & 3394.15 & 0.17 \\
\hline
{Non‐} 
  & {DNA} 
    & Embedding & 0.72 & 284.62 & 0.01\\
 TATA  &BERT-2 & + GC Content & 0.85 &  + 31.25 & 0.02\\
 Promoter & & + zCurve & 0.85 &  + 31.22 & 0.02\\
  & & + AT/GC Ratio & 0.85 &  + 31.17 & 0.02\\
  & & + Cumulative Skew & 0.84 &  + 31.22 & 0.02\\
  & & + PseudoKNC & 0.85 &  + 84.30 & 0.02 \\
  & & Fine‐tuned & \textbf{0.89} & 8886.68 & 0.44 \\
  \cline{2-6}
  & {Hyena} 
    & Embedding & \textbf{0.84} & \textbf{125.41} & 0.01\\
  &DNA & + GC Content & 0.66 &  + 30.88 & 0.01\\
  & & + zCurve & 0.75 &  + 30.68 & 0.01\\
  & & + AT/GC Ratio & 0.83 &  + 31.63 & 0.01\\
  & & + Cumulative Skew & 0.72 &  + 29.98 & 0.01\\
  & & + PseudoKNC & 0.79 &  + 26.12 & 0.01 \\
  & & Fine‐tuned & 0.77 & 940.51 & 0.04 \\
\hline
\end{tabular}
\end{table}

\subsection{Accuracy - Efficiency Trade‐offs in Enhancer and Promoter Classification}
\label{subsec:discussion_tasks}

Table~\ref{tab:merged_classification_results} presents accuracy and inference‐time results for two independent tasks, enhancer classification and non‐TATA promoter classification comparing DNABERT‐2 and Hyenadna embeddings (with and without feature augmentations) against a fully fine‐tuned DNABERT‐2 and Hyenadna model. Several key insights emerge as follows:

\subsubsection{Enhancer Classification}
Embedding-only DNABERT-2 achieves 0.65 accuracy in 525.74 seconds, with carbon emissions of only 0.02 kg. Simple feature augmentations such as GC content, z-curve, AT/GC ratio, and cumulative skew yield modest improvements (up to 0.67 accuracy) for minimal computational overhead (2.24 - 2.61 seconds). For example, adding the AT/GC ratio boosts accuracy to 0.67 with just 2.24 additional seconds. Even the higher-dimensional PseudoKNC feature raises accuracy to 0.67 at the cost of 10.97 seconds. In contrast, full fine-tuning of DNABERT-2 leads to lower accuracy (0.61) and significantly greater compute time ( 74.90 seconds for pre-processing, 31411.48 seconds for fine-tuning and 10.70 seconds for inference), with 1.55 kg of carbon emission.

HyenaDNA embeddings perform comparably, achieving 0.65 accuracy in 410.96 seconds with 0.02 kg carbon emission. zCurve augmentation brings the highest accuracy (0.68) with only 1.94 seconds overhead. However, fine-tuning HyenaDNA yields lower performance (0.58 accuracy) despite 3394.15 (179.98 for pre-processing, 3204.06 for fine-tuning and 10.11 for inference) seconds of compute and 0.17 kg CO$_2$. This highlights the superiority of fixed embedding approaches with simple features in both efficiency and accuracy.

\paragraph{Non‐TATA Promoter Classification.} 
In the non-TATA promoter classification task, DNABERT-2 embeddings combined with handcrafted features offer significant reductions in inference time and carbon emission compared to the fine-tuned model. For instance, DNABERT-2 + AT/GC ratio achieves competitive performance with an inference time of only 315.79 seconds and 0.02 kg CO\textsubscript{2}. In contrast, fine-tuned DNABERT-2 takes 8886.68 seconds and emits 0.44 kg CO\textsubscript{2}.

 On the other hand, the HyenaDNA embedding with AT/GC ratio achieved strong accuracy (0.83) while requiring only 157.04 seconds and emitting 0.01 kg CO\textsubscript{2}. In contrast, the fine-tuned HyenaDNA model required 940.51 seconds and emitted 0.04 kg CO\textsubscript{2}.

\begin{figure}[htbp]
    \centering
    \begin{subfigure}
        \centering
        \includegraphics[width=\textwidth]{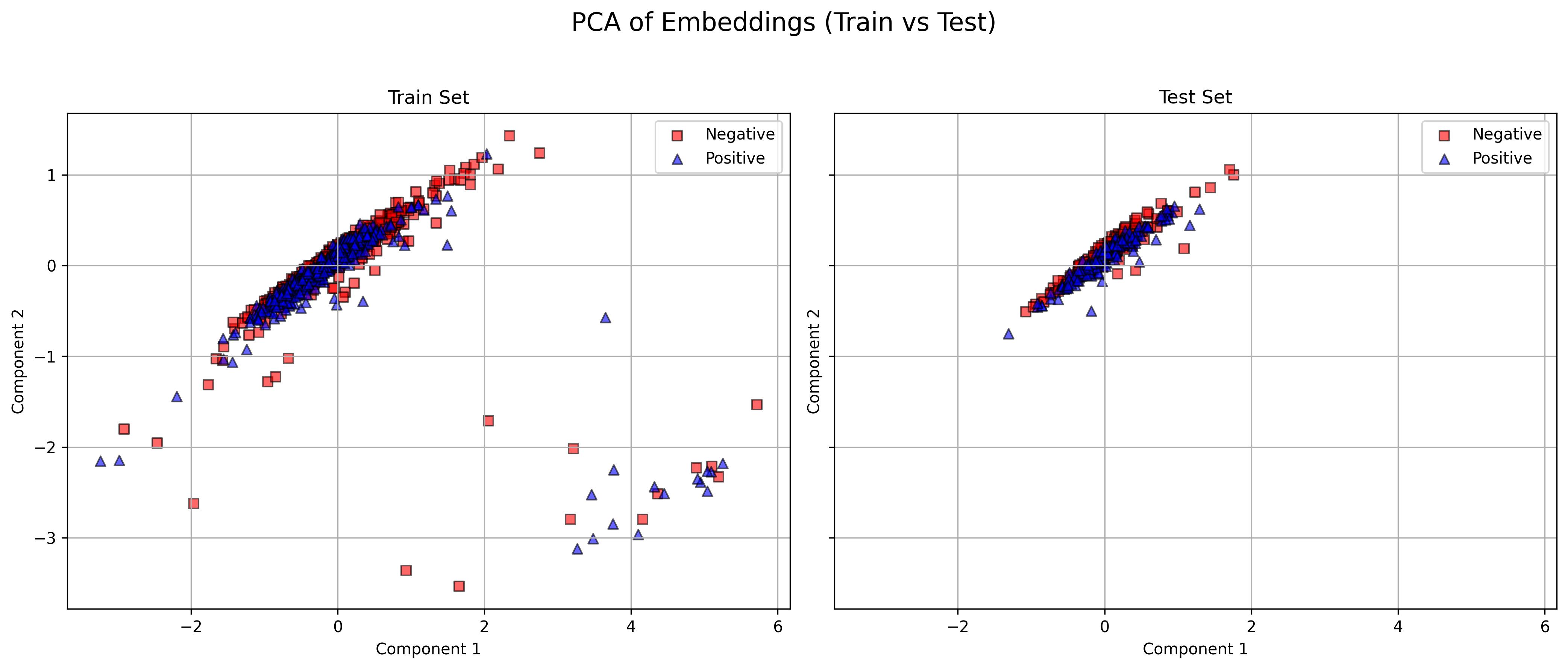}
        \caption{Comparison of dimensionality reduction methods applied to DNABERT-2 sequence embeddings for the enhancer classification task. Both methods show class separation and generalization between training and test sets.}
        \label{images/pca_train_test_side_by_side_enhancer_pca.jpg}
    \end{subfigure}
    \hfill
    \begin{subfigure}
        \centering
        \includegraphics[width=\textwidth]{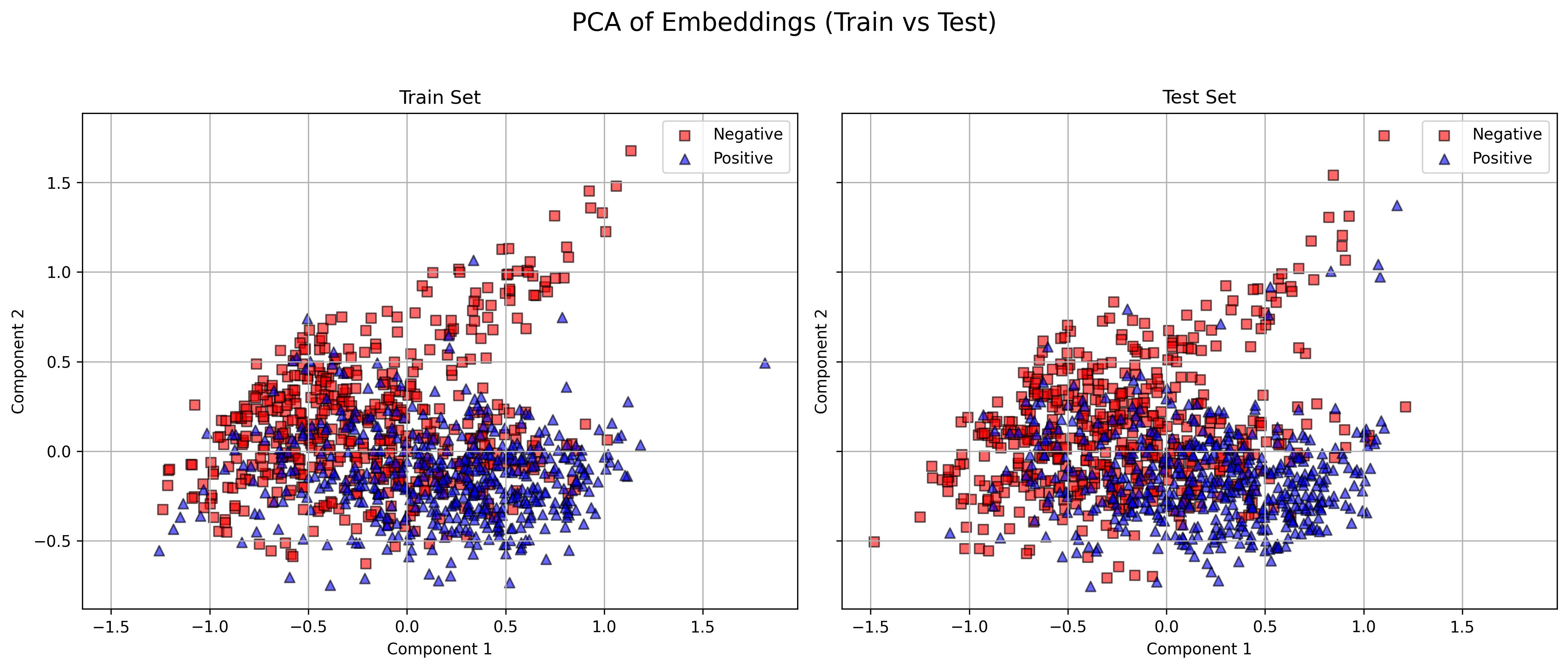}
        
        \label{fig:pca_embeddings}
    \end{subfigure}
    \caption{Comparison of dimensionality reduction methods applied to DNABERT-2 sequence embeddings for the promoter classification task. Both methods show class separation and generalization between training and test sets.}
    \label{fig:embedding_visualizations}
\end{figure}

To further analyze the embedding space learned by the DNABERT-2 model, we applied Principal Component Analysis (PCA) to reduce the high-dimensional embeddings to two dimensions for visualization. We randomly sampled up to 500 positive and 500 negative sequences from both the training and test sets and projected their embeddings into a 2D space using PCA. The resulting plots Figure \ref{images/pca_train_test_side_by_side_enhancer_pca.jpg} and Figure \ref{fig:embedding_visualizations} provide insight into how well the embeddings separate classes and generalize across datasets. Clear separation between positive and negative samples in both sets suggests that the model captures class-discriminative information, while the structural similarity between the training and test embeddings indicates good generalization performance.

% \paragraph{Insights.} 
% Across both tasks, retrieval‐based classification on fixed DNABERT‐2 embeddings (with lightweight feature fusion) consistently matches or exceeds the accuracy of a fully fine‐tuned transformer, while reducing end‐to‐end runtime by over 80\%. In enhancer classification where fixed embeddings alone underperform relative to feature‐augmented variants adding z‐curve or GC content yields substantial accuracy improvements (0.65→0.71) at minimal cost (2–3\,s). In promoter classification where embeddings are already strong feature fusion does not improve accuracy, suggesting that DNABERT‐2 embeddings already capture the necessary sequence signals. Crucially, fine‐tuning DNABERT‐2 is both slower and, in the enhancer task, less accurate than embedding‐based retrieval, underscoring the practical and ecological benefits of our hybrid, retrieval‐augmented pipeline.

\subsection{Carbon Footprint and Efficiency Analysis}
To assess the sustainability and efficiency of the models used in our experiments, we evaluated not only the classification accuracy and the inference time but also estimated carbon emissions. Table \ref{tab:merged_classification_results} includes carbon emission results for both the Enhancer Classification and Non-TATA Promoter Classification tasks. Experiments were conducted using a private infrastructure, which has a carbon efficiency of 0.593 kgCO$_2$eq/kWh. Estimations were conducted using the \href{https://mlco2.github.io/impact#compute}{MachineLearning Impact calculator} presented in \cite{lacoste2019quantifying}.
    
%Uncomment if you bought additional offsets:
%XX kg CO2eq were manually offset through \href{link}{Offset Provider}.

Notably, although fine-tuning DNABERT-2 leads to higher resource consumption and CO$_2$ emissions, the classification accuracy does not consistently improve over the embedding-based approaches with auxiliary features. This insight emphasizes the trade-off between accuracy, computational cost, and environmental impact, advocating for green AI practices in bioinformatics model development.
\section{Discussion}

\subsection{Performance vs. Carbon Efficiency}

We conducted a comprehensive evaluation of DNABERT-2 and its variants by augmenting the base embeddings with biologically inspired handcrafted features, and compared these against fine-tuned DNABERT-2 models across two downstream tasks: {Enhancer Classification} and {Non-TATA Promoter Classification}.
\paragraph{Embedding vs. Fine-Tuning.}
As shown in Table~\ref{tab:merged_classification_results}, embedding-based approaches closely match or outperform fine-tuned models with significantly lower inference time and carbon emissions. For enhancer classification, DNABERT-2 with zCurve achieved an accuracy of 0.67 (vs. 0.61 for fine-tuning), while reducing emissions from 1.55 kg to 0.02 kg CO\textsubscript{2} and inference time from over 31,000 s to ~528 s. HyenaDNA with zCurve slightly improved accuracy to 0.68 and brought inference time to just 412.9 s and emissions to 0.02 kg.

In the non-TATA promoter classification task, DNABERT-2 with zCurve achieved 0.85 accuracy close to the fine-tuned model’s 0.89 while cutting emissions from 0.44 kg to 0.02 kg CO\textsubscript{2} and runtime from 8,886s to ~316s. HyenaDNA’s zCurve variant reached 0.75 accuracy with just 156s of total inference time and 0.01kg CO\textsubscript{2}, offering high efficiency at modest accuracy trade-offs.

\subsection{Impact of Handcrafted Features}
Augmenting transformer embeddings with handcrafted features improved classification in most cases. For enhancer prediction, adding zCurve to DNABERT-2 increased accuracy from 0.65 to 0.67, and from 0.65 to 0.68 with HyenaDNA. Features like AT/GC ratio and cumulative skew also yielded moderate gains. However, PseudoKNC, despite being computationally intensive, offered no significant accuracy advantage, suggesting limited benefits from high-dimensional features in this context.

\subsection{Generalization Across Datasets}

Table~\ref{tab:full_results} demonstrates the generalization capabilities of embedding-based models over nine genomic benchmarks. DNABERT-2 embeddings combined with statistical features reached top-1\% accuracies up to {0.91} (Human vs. Worm) and showed consistent improvement on challenging datasets like {Mouse Enhancers} and {Human OCR Ensembl}. Importantly, these results were obtained {without fine-tuning}, confirming the utility of pre-trained DNA embeddings for downstream tasks.

\subsection{Environmental Implications}

The observed {up to 77.5$\times$ reduction in CO\textsubscript{2}} emissions makes embedding-based workflows far more sustainable and aligns with {Green AI principles}. Fine-tuning large genomic models requires substantial compute and energy, limiting practical deployment. In contrast, our embedding-based pipeline offers a scalable, low-emission alternative for real-world applications where environmental cost is a concern. These observations align with \emph{Green AI} principles, advocating for environmentally sustainable methods that balance accuracy with compute cost \cite{schwartz2020green,patterson2021carbon}.

\section{Conclusion}
\label{sec:conclusion}

In this study, we investigated the efficacy and efficiency of leveraging embeddings combined with biologically inspired handcrafted features for genomic sequence classification. Our results demonstrate that embedding-based approaches can achieve comparable or superior performance to fine-tuned transformer models, while drastically reducing inference time and carbon emissions achieving up to 77.5× lower CO\textsubscript{2} emissions in certain scenarios.

Despite these benefits, our approach has some limitations. Handcrafted features are fixed and may not capture task-specific contextual nuances, and static pretrained embeddings lack adaptability during training. However, this deliberate design enables significant reductions in computational cost and carbon footprint, without compromising accuracy aligning with the goals of sustainable and efficient AI research in genomics.

For future work, we plan to investigate hybrid models that integrate task-specific fine-tuning with fixed embeddings to balance adaptability and efficiency. We also aim to extend this framework to more complex regulatory genomics tasks, including enhancer-promoter interaction prediction and multi-label regulatory element classification.

\bibliographystyle{unsrt}  
\bibliography{references}  %%% Remove comment to use the external .bib file (using bibtex).


% George Kour and Raid Saabne.
% \newblock Real-time segmentation of on-line handwritten arabic script.
% \newblock In {\em Frontiers in Handwriting Recognition (ICFHR), 2014 14th
%   International Conference on}, pages 417--422. IEEE, 2014.

% George Kour and Raid Saabne.
% \newblock Fast classification of handwritten on-line arabic characters.
% \newblock In {\em Soft Computing and Pattern Recognition (SoCPaR), 2014 6th
%   International Conference of}, pages 312--318. IEEE, 2014.

% \bibitem{hadash2018estimate}
% Guy Hadash, Einat Kermany, Boaz Carmeli, Ofer Lavi, George Kour, and Alon
% \newblock Estimate and replace: A novel approach to integrating deep neural
%   networks with existing applications.
% \newblock {\em arXiv preprint arXiv:1804.09028}, 2018.

% \end{thebibliography}
\newpage
\section{Appendix}

\subsubsection{Handcrafted Feature Details}

% Inspired by PyFeat \cite{muhammod2019pyfeat}, to complement the high-dimensional representations from pretrained embeddings, we extract following set of biologically-inspired handcrafted features that capture essential patterns and characteristics of DNA sequences .

\begin{enumerate}
 \item \textbf{z-Curve Representation.}
The z-Curve is a three-dimensional curve that uniquely maps a nucleotide sequence based on nucleotide disparity in three orthogonal directions- 
 \begin{align}
    x &= (A + G) - (C + T),\\ 
    y &= (A + C) - (G + T) \text{ and}, \\
    z &= (A + T) - (G + C),
 \end{align}
For rest of the sections, $A$, $T$, $G$, and $C$ are the total counts of each base in the sequence. These components capture purine-pyrimidine imbalance ($x$), amino-keto group content ($y$), and weak-strong hydrogen bonding asymmetry ($z$).

 \item \textbf{GC Content.}
GC content reflects the proportion of guanine and cytosine in the sequence and is defined as:
\begin{equation}
    \text{GC content} = \frac{G + C}{A + T + G + C} \times 100\%.
\end{equation}

 \item \textbf{AT/GC Ratio.}
The AT/GC ratio indicates compositional bias and is computed as:
\begin{equation}
    \text{AT/GC ratio} = \frac{A + T}{G + C}.
\end{equation}

 \item \textbf{Cumulative Skews.}
Nucleotide skews are widely used to detect strand asymmetries. We compute:\\
\[
\text{GC skew} = \frac{G - C}{G + C}, \quad \text{AT skew} = \frac{A - T}{A + T}
\]

 \item \textbf{Pseudo k-tuple Nucleotide Composition (PseKNC).}
To capture both the local composition and sequence-order information, we use PseKNC. For a fixed $k$ (typically $k=3$), all possible $k$-mers from the nucleotide alphabet $\{A, C, G, T\}$ are enumerated. The number of possible $k$-mers \text{\# k-mers} = 4$^k$.

% When $k=3$, this results in $64$ basic features. Additional correlation factors can be included, yielding a total of $84$ features. PseKNC can also be extended to proteins using the 20 amino acid alphabet, producing $20^3 = 8000$ base features plus correlation features, totaling $8420$.
    % \item \textbf{AT/GC Ratio}: Measures the proportion of adenine-thymine (A-T) to guanine-cytosine (G-C) nucleotides:
    % \[
    % \text{AT/GC Ratio} = \frac{\text{Count}(A) + \text{Count}(T)}{\text{Count}(G) + \text{Count}(C) + \epsilon}
    % \]
    % where $\epsilon$ is a small constant to avoid division by zero.

    % \item \textbf{Z-Curve Features}: A 3-dimensional curve representation of DNA sequence composition:
    % \[
    % x(n) = (A + G) - (C + T), \quad y(n) = (A + C) - (G + T), \quad z(n) = (A + T) - (G + C)
    % \]
    % These features capture nucleotide imbalance and distribution patterns.

    % \item \textbf{GC and AT Skew}: Reflects strand asymmetry and mutational bias:
    % \[
    % \text{GC Skew} = \frac{G - C}{G + C}, \quad \text{AT Skew} = \frac{A - T}{A + T}
    % \]

    % \item \textbf{K-mer Composition (k=3)}: Counts the frequency of all possible 3-mers (triplets) to represent local sequence context. For DNA, there are $4^3 = 64$ possible triplets. Frequencies are normalized by sequence length.

    % \item \textbf{Pseudo k-tuple Nucleotide Composition (PseKNC)}: Extends k-mer composition by incorporating the physicochemical properties of dinucleotides and higher-order motifs to capture sequence order effects.

\end{enumerate}
% \newpage
% \onecolumn
% \begin{figure*}[ht]
%     \centering
%     \includegraphics[width=0.55\textwidth]{images/t-sne all updated new.jpg}
%     \caption{
%         t-SNE visualization of DNABERT2 embeddings across all nine genomic benchmark datasets. 
%         We randomly sampled 100 sequences from each dataset to ensure balanced comparison. 
%         Each point represents a high-dimensional embedding projected into two dimensions using t-SNE. 
%         Colors denote different datasets, while marker shapes correspond to different target classes within each dataset. 
%         The visualization highlights how DNABERT2 captures dataset-specific and task-relevant features in the embedding space. The numbering on the right side of the plot follows the mapping of task number given in column \# presented in Table \ref{tab:dataset_summary}.
%     }
%     \label{fig:tsne_dnabert}
% \end{figure*}

\newpage
\onecolumn
\subsection{Algorithm}
\label{algo}
\begin{algorithm}[H]
\caption{Retrieval‐Augmented Genome Classification (Transformer‐Agnostic)}
\label{alg:rag_generic}
\begin{algorithmic}[1]
\Require 
\Statex Training set $\mathcal{D}_{\text{train}} = \{(\mathbf{x}_i, y_i)\}_{i=1}^N$, \\
\quad\enspace Test set $\mathcal{D}_{\text{test}} = \{\mathbf{x}_j\}_{j=1}^M$, \\
\quad\enspace Pretrained transformer encoder $f_{\theta}$, \\
\quad\enspace Number of neighbors $k$.
\Ensure Predicted labels $\{\hat{y}_j\}_{j=1}^M$ and evaluation metrics.

\vspace{0.5em}
\State \textbf{Embedding Phase (offline):}
\ForAll{$(\mathbf{x}_i, y_i)\in \mathcal{D}_{\text{train}}$} 
    \Comment{Compute embedding for each training sequence}
    \State $\mathbf{e}_i \gets$ \textsc{EmbedSequence}($f_{\theta}, \mathbf{x}_i$) 
    \Comment{e.g., mean‐pool transformer outputs}
\EndFor
\State Form matrix $E_{\text{train}} \in \mathbb{R}^{N\times D}$ with rows $\mathbf{e}_i$.

\vspace{0.5em}
\State \textbf{Index Construction:}
\State Build FAISS $L_2$ index $\mathcal{I} \gets \text{IndexFlatL2}(D)$.
\State $\mathcal{I}.\text{add}(E_{\text{train}})$  \Comment{Add all training embeddings}

\vspace{0.5em}
\State \textbf{Inference Phase (online):}
\State Start timer $t_{\text{start}} \gets \text{now}()$.

\ForAll{$\mathbf{x}_j \in \mathcal{D}_{\text{test}}$}
    \Comment{Embed each test sequence}
    \State $\mathbf{e}_j \gets$ \textsc{EmbedSequence}($f_{\theta}, \mathbf{x}_j$)
    \Comment{Obtain $D$‐dimensional vector}
    
    \State $[\,\_,\, i_{j,1},\ldots,i_{j,k}\,] \gets \mathcal{I}.\text{search}(\mathbf{e}_j, k)$  
    \Comment{Retrieve indices of $k$ nearest neighbors}
    
    \State $\{y_{i_{j,1}},\dots,y_{i_{j,k}}\} \gets$ labels of retrieved neighbors
    \Comment{Lookup ground‐truth labels in training set}
    
    \State $\hat{y}_j \gets \arg\max_{c} \sum_{m=1}^k \mathbf{1}\bigl(y_{i_{j,m}}=c\bigr)$  
    \Comment{Majority‐voting across neighbor labels}
\EndFor

\State Stop timer $t_{\text{end}} \gets \text{now}()$.

\vspace{0.5em}
\State \textbf{Evaluation:}
\State Compute accuracy, F1, and other metrics comparing $\{\hat{y}_j\}$ to true labels.
\State Report wall‐clock time: $\Delta t = t_{\text{end}} - t_{\text{start}}$.

\Function{EmbedSequence}{$f_{\theta}, \mathbf{x}$}
    \Comment{Return $D$‐dimensional mean‐pooled embedding}
    \State $\text{tokens} \gets \text{Tokenize}(\mathbf{x})$
    \State $\text{hidden} \gets f_{\theta}(\text{tokens})$  
    \Comment{Output shape: $(L, D)$ for $L$ tokens}
    \State \Return $\frac{1}{L} \sum_{t=1}^{L} \text{hidden}_t$
\EndFunction

\end{algorithmic}
\end{algorithm}
Algorithm~\ref{alg:rag_generic} presents a transformer-agnostic retrieval-augmented classification framework for genomic sequences. 

In the \textbf{embedding phase} (performed offline), each training sequence $\mathbf{x}_i$ is processed by a pretrained transformer encoder $f_{\theta}$ to extract a fixed-length embedding vector $\mathbf{e}_i$, obtained by mean pooling over the transformer's token-level outputs. These embeddings are collected into a matrix $E_{\text{train}}$.

Next, an efficient nearest neighbor index is built over the training embeddings using FAISS with an $L_2$ distance metric to enable fast similarity search.

During the \textbf{inference phase} (online), each test sequence $\mathbf{x}_j$ is similarly embedded into vector $\mathbf{e}_j$. The FAISS index is queried to retrieve the indices of the $k$ nearest training neighbors to $\mathbf{e}_j$. The predicted label $\hat{y}_j$ for the test sequence is obtained by majority voting over the labels of the retrieved neighbors.

Finally, standard classification metrics such as accuracy and F1-score are computed by comparing the predicted labels $\{\hat{y}_j\}$ to the true test labels. The total inference runtime is also recorded to evaluate computational efficiency.

This approach enables flexible integration with any pretrained transformer encoder and leverages similarity search to perform non-parametric classification based on learned embeddings.

\begin{algorithm}[H]
\caption{Part 1: Feature Extraction and Hybrid Embedding Construction}
\label{alg:feature_extraction}
\begin{algorithmic}[1]
\Require Training set $\mathcal{D}_{\text{train}}$, Test set $\mathcal{D}_{\text{test}}$, embedding vectors $\{\mathbf{e}_i\}$, $\{\mathbf{e}_j\}$, feature functions $\mathcal{F}$
\Ensure Hybrid embeddings for train/test: $\{\mathbf{h}_{i,\text{train}}\}$, $\{\mathbf{h}_{j,\text{test}}\}$

\ForAll{$(\mathbf{x}_i, y_i) \in \mathcal{D}_{\text{train}}$}
    \State Initialize $\mathbf{h}_{i,\text{feat}} \gets []$
    \ForAll{$f \in \mathcal{F}$}
        \State $v \gets f(\mathbf{x}_i)$
        \If{$v$ scalar} \State $v \gets [v]$
        \EndIf
        \State Append $v$ to $\mathbf{h}_{i,\text{feat}}$
    \EndFor
    \State $\mathbf{h}_{i,\text{feat}} \gets \text{MinMaxScale}(\mathbf{h}_{i,\text{feat}})$
    \State $\tilde{\mathbf{e}}_i \gets \mathbf{e}_i / \|\mathbf{e}_i\|$
    \State $\mathbf{h}_{i,\text{train}} \gets [\,\alpha\,\tilde{\mathbf{e}}_i \,\|\, \beta\,\mathbf{h}_{i,\text{feat}}\,]$
\EndFor

\ForAll{$\mathbf{x}_j \in \mathcal{D}_{\text{test}}$}
    \State Compute $\mathbf{h}_{j,\text{feat}}$ similarly
    \State $\tilde{\mathbf{e}}_j \gets \mathbf{e}_j / \|\mathbf{e}_j\|$
    \State $\mathbf{h}_{j,\text{test}} \gets [\,\alpha\,\tilde{\mathbf{e}}_j \,\|\, \beta\,\mathbf{h}_{j,\text{feat}}\,]$
\EndFor
\end{algorithmic}
\end{algorithm}
\begin{algorithm}[H]
\caption{Part 2: FAISS Indexing and Dynamic k-NN Retrieval}
\label{alg:dynamic_retrieval}
\begin{algorithmic}[1]
\Require Hybrid embeddings $\{\mathbf{h}_{i,\text{train}}\}$, $\{\mathbf{h}_{j,\text{test}}\}$, base $k_{\text{base}}$, max factor $\alpha_{\max}$, threshold $\tau$
\Ensure Candidate neighbors and similarities for each test point

\State Build FAISS index $\mathcal{I} \gets \text{IndexFlatIP}(\cdot)$
\State $\mathcal{I}.\text{add}(\{\mathbf{h}_{i,\text{train}}\})$
\State $k_{\max} \gets \min(\alpha_{\max} \cdot k_{\text{base}}, N)$
\State $(D, I) \gets \mathcal{I}.\text{search}(\{\mathbf{h}_{j,\text{test}}\}, k_{\max})$

\For{$j = 1$ to $M$}
    \State $\mu_j \gets \frac{1}{k_{\text{base}}} \sum_{m=1}^{k_{\text{base}}} D_{j,m}$
    \If{$\mu_j < \tau$}
        \State Use $k_{\max}$ neighbors: $D_j \gets D_{j,1:k_{\max}},\; I_j \gets I_{j,1:k_{\max}}$
    \Else
        \State Use $k_{\text{base}}$ neighbors: $D_j \gets D_{j,1:k_{\text{base}}},\; I_j \gets I_{j,1:k_{\text{base}}}$
    \EndIf
\EndFor
\end{algorithmic}
\end{algorithm}
\begin{algorithm}[H]
\caption{Part 3: Weighted Voting and Final Prediction}
\label{alg:weighted_voting}
\begin{algorithmic}[1]
\Require Neighbor sets $\{I_j\}$, distances $\{D_j\}$, class labels of training set
\Ensure Predicted labels $\{\hat{y}_j\}$

\For{$j = 1$ to $M$}
    \State Initialize vote vector $\mathbf{v}_j \gets \mathbf{0} \in \mathbb{R}^C$
    \For{$m = 1$ to $|I_j|$}
        \State $w_{j,m} \gets \frac{1}{1 + e^{-D_{j,m}}}$ \Comment{Sigmoid weighting}
        \State $c \gets$ class label of training index $I_j[m]$
        \State $\mathbf{v}_j[c] \gets \mathbf{v}_j[c] + w_{j,m}$
    \EndFor
    \State $\hat{y}_j \gets \arg\max_c \mathbf{v}_j[c]$
\EndFor

\State Compute accuracy, F1-score, and runtime
\end{algorithmic}
\end{algorithm}
The overall classification procedure along with handcrafted features is divided into three main stages, described in Algorithms~\ref{alg:feature_extraction}, \ref{alg:dynamic_retrieval}, and \ref{alg:weighted_voting}.

\textbf{Algorithm~\ref{alg:feature_extraction}} outlines the \textit{Feature Extraction and Hybrid Embedding Construction} step. Here, handcrafted features are computed from input sequences using a predefined set of functions, scaled via min-max normalization, and concatenated with normalized pretrained embeddings to form hybrid feature vectors for both training and test samples.

\textbf{Algorithm~\ref{alg:dynamic_retrieval}} describes the \textit{FAISS Indexing and Dynamic k-NN Retrieval} phase. We build an approximate nearest neighbor index using the training hybrid embeddings and perform batched retrieval for each test embedding. The number of neighbors considered is dynamically adjusted based on a similarity threshold, enabling adaptive neighbor selection.

Finally, \textbf{Algorithm~\ref{alg:weighted_voting}} presents the \textit{Weighted Voting and Final Prediction} step. Retrieved neighbors vote for the class label of each test sample, with weights computed by applying a sigmoid function to similarity scores. The predicted label corresponds to the class with the highest cumulative weighted vote. Performance metrics such as accuracy and F1-score are then computed to evaluate classification effectiveness.

\section*{Detailed Classification Reports on Genomic Benchmark Datasets (Embedding only) }

% ---------------- DNABERT-2 ------------------
\subsection*{DNABERT-2}

\subsubsection*{Coding vs Intergenomic Seqs}
\begin{tabular}{lcccc}
\toprule
Class & Precision & Recall & F1-score & Support \\
\midrule
0 & 0.81 & 0.93 & 0.86 & 12500 \\
1 & 0.91 & 0.78 & 0.84 & 12500 \\
\midrule
Accuracy &       &       & \textbf{0.85} & 25000 \\

\bottomrule
\end{tabular}

\vspace{1em}

\subsubsection*{Human or Worm}
\begin{tabular}{lcccc}
\toprule
Class & Precision & Recall & F1-score & Support \\
\midrule
0 & 0.95 & 0.87 & 0.91 & 12500 \\
1 & 0.88 & 0.96 & 0.92 & 12500 \\
\midrule
Accuracy &       &       & \textbf{0.91} & 25000 \\

\bottomrule
\end{tabular}

\vspace{1em}

\subsubsection*{Drosophila Enhancers Stark}
\begin{tabular}{lcccc}
\toprule
Class & Precision & Recall & F1-score & Support \\
\midrule
0 & 0.73 & 0.63 & 0.67 & 865 \\
1 & 0.67 & 0.77 & 0.72 & 865 \\
\midrule
Accuracy &       &       & \textbf{0.70} & 1730 \\
\bottomrule
\end{tabular}

\vspace{1em}

\subsubsection*{Mouse Enhancers Ensembl}
\begin{tabular}{lcccc}
\toprule
Class & Precision & Recall & F1-score & Support \\
\midrule
0 & 0.77 & 0.61 & 0.68 & 121 \\
1 & 0.68 & 0.82 & 0.74 & 121 \\
\midrule
Accuracy &       &       & \textbf{0.71} & 242 \\
\bottomrule
\end{tabular}

\vspace{1em}

\subsubsection*{Human Enhancers Cohn}
\begin{tabular}{lcccc}
\toprule
Class & Precision & Recall & F1-score & Support \\
\midrule
0 & 0.71 & 0.68 & 0.70 & 3474 \\
1 & 0.69 & 0.72 & 0.71 & 3474 \\
\midrule
Accuracy &       &       & \textbf{0.70} & 6948 \\
\bottomrule
\end{tabular}

\vspace{1em}

\subsubsection*{Human Enhancers Ensembl}
\begin{tabular}{lcccc}
\toprule
Class & Precision & Recall & F1-score & Support \\
\midrule
0 & 0.75 & 0.65 & 0.69 & 15485 \\
1 & 0.69 & 0.78 & 0.73 & 15485 \\
\midrule
Accuracy &       &       & \textbf{0.71} & 30970 \\
\bottomrule
\end{tabular}

\vspace{1em}

\subsubsection*{Human Ensembl Regulatory}
\begin{tabular}{lcccc}
\toprule
Class & Precision & Recall & F1-score & Support \\
\midrule
0 & 0.50 & 0.50 & 0.50 & 21378 \\
1 & 0.50 & 0.60 & 0.54 & 17476 \\
2 & 0.81 & 0.67 & 0.73 & 18859 \\
\midrule
Accuracy &       &       & \textbf{0.58} & 57713 \\
\bottomrule
\end{tabular}

\vspace{1em}

\subsubsection*{Human OCR Ensembl}
\begin{tabular}{lcccc}
\toprule
Class & Precision & Recall & F1-score & Support \\
\midrule
0 & 0.65 & 0.59 & 0.62 & 17476 \\
1 & 0.62 & 0.68 & 0.65 & 17476 \\
\midrule
Accuracy &       &       & \textbf{0.63} & 34952 \\
\bottomrule
\end{tabular}

\vspace{1em}

\subsubsection*{Human Non-TATA Promoters}
\begin{tabular}{lcccc}
\toprule
Class & Precision & Recall & F1-score & Support \\
\midrule
0 & 0.80 & 0.99 & 0.88 & 4119 \\
1 & 0.99 & 0.79 & 0.88 & 4915 \\
\midrule
Accuracy &       &       & \textbf{0.88} & 9034 \\
\bottomrule
\end{tabular}
\subsection*{HyenaDNA}

\subsubsection*{Coding vs Intergenomic Seqs}
\begin{tabular}{lcccc}
\toprule
Class & Precision & Recall & F1-score & Support \\
\midrule
0 & 0.81 & 0.85 & 0.83 & 12500 \\
1 & 0.84 & 0.80 & 0.82 & 12500 \\
\midrule
Accuracy &       &       & \textbf{0.83} & 25000 \\
\bottomrule
\end{tabular}

\vspace{1em}

\subsubsection*{Human or Worm}
\begin{tabular}{lcccc}
\toprule
Class & Precision & Recall & F1-score & Support \\
\midrule
0 & 0.82 & 0.84 & 0.83 & 12500 \\
1 & 0.84 & 0.82 & 0.83 & 12500 \\
\midrule
Accuracy &       &       & \textbf{0.83} & 25000 \\
\bottomrule
\end{tabular}

\vspace{1em}

\subsubsection*{Drosophila Enhancers Stark}
\begin{tabular}{lcccc}
\toprule
Class & Precision & Recall & F1-score & Support \\
\midrule
0 & 0.63 & 0.52 & 0.57 & 865 \\
1 & 0.59 & 0.70 & 0.64 & 865 \\
\midrule
Accuracy &       &       & \textbf{0.61} & 1730 \\
\bottomrule
\end{tabular}

\vspace{1em}

\subsubsection*{Mouse Enhancers Ensembl}
\begin{tabular}{lcccc}
\toprule
Class & Precision & Recall & F1-score & Support \\
\midrule
0 & 0.74 & 0.72 & 0.73 & 121 \\
1 & 0.73 & 0.74 & 0.73 & 121 \\
\midrule
Accuracy &       &       & \textbf{0.73} & 242 \\
\bottomrule
\end{tabular}

\vspace{1em}

\subsubsection*{Human Enhancers Ensembl}
\begin{tabular}{lcccc}
\toprule
Class & Precision & Recall & F1-score & Support \\
\midrule
0 & 0.72 & 0.66 & 0.69 & 15485 \\
1 & 0.69 & 0.75 & 0.72 & 15485 \\
\midrule
Accuracy &       &       & \textbf{0.70} & 30970 \\
\bottomrule
\end{tabular}

\vspace{1em}

\subsubsection*{Human Enhancers Cohn}
\begin{tabular}{lcccc}
\toprule
Class & Precision & Recall & F1-score & Support \\
\midrule
0 & 0.68 & 0.65 & 0.66 & 3474 \\
1 & 0.66 & 0.69 & 0.68 & 3474 \\
\midrule
Accuracy &       &       & \textbf{0.67} & 6948 \\
\bottomrule
\end{tabular}

\vspace{1em}

\subsubsection*{Human Ensembl Regulatory}
\begin{tabular}{lcccc}
\toprule
Class & Precision & Recall & F1-score & Support \\
\midrule
0 & 0.69 & 0.86 & 0.77 & 21378 \\
1 & 0.87 & 0.72 & 0.79 & 17476 \\
2 & 0.84 & 0.73 & 0.78 & 18859 \\
\midrule
Accuracy &       &       & \textbf{0.78} & 57713 \\
\bottomrule
\end{tabular}

\vspace{1em}

\subsubsection*{Human OCR Ensembl}
\begin{tabular}{lcccc}
\toprule
Class & Precision & Recall & F1-score & Support \\
\midrule
0 & 0.63 & 0.58 & 0.60 & 17476 \\
1 & 0.61 & 0.66 & 0.63 & 17476 \\
\midrule
Accuracy &       &       & \textbf{0.62} & 34952 \\
\bottomrule
\end{tabular}

\vspace{1em}

\subsubsection*{Human Non-TATA Promoters}
\begin{tabular}{lcccc}
\toprule
Class & Precision & Recall & F1-score & Support \\
\midrule
0 & 0.75 & 0.89 & 0.81 & 4119 \\
1 & 0.89 & 0.76 & 0.82 & 4915 \\
\midrule
Accuracy &       &       & \textbf{0.82} & 9034 \\
\bottomrule
\end{tabular}

% ---------------- Nucleotide Transformer ------------------

\subsection*{Nucleotide Transformer}

\subsubsection*{Human or Worm}
\begin{tabular}{lcccc}
\toprule
Class & Precision & Recall & F1-score & Support \\
\midrule
0 & 0.89 & 0.92 & 0.90 & 12500 \\
1 & 0.92 & 0.88 & 0.90 & 12500 \\
\midrule
Accuracy &       &       & \textbf{0.90} & 25000 \\
\bottomrule
\end{tabular}

\vspace{1em}

\subsubsection*{Coding vs Intergenomic Seqs}
\begin{tabular}{lcccc}
\toprule
Class & Precision & Recall & F1-score & Support \\
\midrule
0 & 0.90 & 0.80 & 0.85 & 12500 \\
1 & 0.82 & 0.91 & 0.86 & 12500 \\
\midrule
Accuracy &       &       & \textbf{0.85} & 25000 \\
\bottomrule
\end{tabular}

\vspace{1em}

\subsubsection*{Drosophila Enhancers Stark}
\begin{tabular}{lcccc}
\toprule
Class & Precision & Recall & F1-score & Support \\
\midrule
0 & 0.70 & 0.58 & 0.64 & 865 \\
1 & 0.64 & 0.75 & 0.69 & 865 \\
\midrule
Accuracy &       &       & \textbf{0.66} & 1730 \\
\bottomrule
\end{tabular}

\vspace{1em}

\subsubsection*{Mouse Enhancers Ensembl}
\begin{tabular}{lcccc}
\toprule
Class & Precision & Recall & F1-score & Support \\
\midrule
0 & 0.77 & 0.81 & 0.79 & 121 \\
1 & 0.80 & 0.75 & 0.77 & 121 \\
\midrule
Accuracy &       &       & \textbf{0.78} & 242 \\
\bottomrule
\end{tabular}

\vspace{1em}

\subsubsection*{Human Enhancers Cohn}
\begin{tabular}{lcccc}
\toprule
Class & Precision & Recall & F1-score & Support \\
\midrule
0 & 0.66 & 0.66 & 0.66 & 3474 \\
1 & 0.66 & 0.66 & 0.66 & 3474 \\
\midrule
Accuracy &       &       & \textbf{0.66} & 6948 \\
\bottomrule
\end{tabular}

\vspace{1em}

\subsubsection*{Human Ensembl Regulatory}
\begin{tabular}{lcccc}
\toprule
Class & Precision & Recall & F1-score & Support \\
\midrule
0 & 0.49 & 0.59 & 0.54 & 21378 \\
1 & 0.51 & 0.37 & 0.43 & 17476 \\
2 & 0.69 & 0.72 & 0.71 & 18859 \\
\midrule
Accuracy &       &       & \textbf{0.57} & 57713 \\
\bottomrule
\end{tabular}

\vspace{1em}

\subsubsection*{Human Non-TATA Promoters}
\begin{tabular}{lcccc}
\toprule
Class & Precision & Recall & F1-score & Support \\
\midrule
0 & 0.75 & 0.86 & 0.80 & 4119 \\
1 & 0.86 & 0.75 & 0.81 & 4915 \\
\midrule
Accuracy &       &       & \textbf{0.80} & 9034 \\
\bottomrule
\end{tabular}

\vspace{1em}

\subsubsection*{Human Enhancers Ensembl}
\begin{tabular}{lcccc}
\toprule
Class & Precision & Recall & F1-score & Support \\
\midrule
0 & 0.70 & 0.63 & 0.66 & 15485 \\
1 & 0.66 & 0.74 & 0.70 & 15485 \\
\midrule
Accuracy &       &       & \textbf{0.68} & 30970 \\
\end{tabular}
\subsubsection*{Human OCR Ensembl}
\begin{tabular}{lcccc}
\toprule
Class & Precision & Recall & F1-score & Support \\
\midrule
0 & 0.60 & 0.57 & 0.58 & 17476 \\
1 & 0.59 & 0.63 & 0.61 & 17476 \\
\midrule
Accuracy &       &       & \textbf{0.60} & 34952 \\
\bottomrule
\end{tabular}
\newpage
\section*{ Classification Reports on New Independent Test Set}

% ---------------- Enhancer Classification ------------------
\subsection*{Enhancer Classification}

\subsubsection*{HyenaDNA (Embedding only)}
\begin{tabular}{lcccc}
\toprule
Class & Precision & Recall & F1-score & Support \\
\midrule
0 & 0.73 & 0.47 & 0.57 & 200 \\
1 & 0.61 & 0.82 & 0.70 & 200 \\
\midrule
Accuracy &       &       & \textbf{0.65} & 400 \\
\bottomrule
\end{tabular}

\vspace{1em}

\subsubsection*{DNABERT-2 (Embedding only)}
\begin{tabular}{lcccc}
\toprule
Class & Precision & Recall & F1-score & Support \\
\midrule
0 & 0.65 & 0.66 & 0.65 & 200 \\
1 & 0.65 & 0.64 & 0.65 & 200 \\
\midrule
Accuracy &       &       & \textbf{0.65} & 400 \\
\bottomrule
\end{tabular}

\vspace{1em}

\subsubsection*{HyenaDNA (Fine-tuned)}
\begin{tabular}{lcccc}
\toprule
Class & Precision & Recall & F1-score & Support \\
\midrule
0 & 0.56 & 0.77 & 0.65 & 200 \\
1 & 0.63 & 0.40 & 0.49 & 200 \\
\midrule
Accuracy &       &       & \textbf{0.58} & 400 \\
\bottomrule
\end{tabular}

\vspace{1em}

\subsubsection*{DNABERT-2 (Fine-tuned)}
\begin{tabular}{lcccc}
\toprule
Class & Precision & Recall & F1-score & Support \\
\midrule
0 & 0.57 & 0.87 & 0.69 & 200 \\
1 & 0.73 & 0.34 & 0.47 & 200 \\
\midrule
Accuracy &       &       & \textbf{0.61} & 400 \\
\bottomrule
\end{tabular}

% ---------------- Promoter Classification ------------------
\newpage
\subsection*{Promoter Classification}

\subsubsection*{HyenaDNA (Embedding only)}
\begin{tabular}{lcccc}
\toprule
Class & Precision & Recall & F1-score & Support \\
\midrule
0 & 0.86 & 0.87 & 0.87 & 27731 \\
1 & 0.82 & 0.80 & 0.81 & 19811 \\
\midrule
Accuracy &       &       & \textbf{0.84} & 47542 \\
\bottomrule
\end{tabular}

\vspace{1em}

\subsubsection*{DNABERT-2 (Embedding only)}
\begin{tabular}{lcccc}
\toprule
Class & Precision & Recall & F1-score & Support \\
\midrule
0 & 0.95 & 0.54 & 0.69 & 27731 \\
1 & 0.60 & 0.96 & 0.74 & 19811 \\
\midrule
Accuracy &       &       & \textbf{0.72} & 47542 \\
\bottomrule
\end{tabular}

\vspace{1em}

\subsubsection*{DNABERT-2 (Fine-tuned)}
\begin{tabular}{lcccc}
\toprule
Class & Precision & Recall & F1-score & Support \\
\midrule
0 & 0.92 & 0.88 & 0.90 & 27731 \\
1 & 0.85 & 0.90 & 0.87 & 19811 \\
\midrule
Accuracy &       &       & \textbf{0.89} & 47542 \\
\bottomrule
\end{tabular}

\vspace{1em}

\subsubsection*{HyenaDNA (Fine-tuned)}
\begin{tabular}{lcccc}
\toprule
Class & Precision & Recall & F1-score & Support \\
\midrule
0 & 0.80 & 0.82 & 0.81 & 27731 \\
1 & 0.73 & 0.71 & 0.72 & 19811 \\
\midrule
Accuracy &       &       & \textbf{0.77} & 47542 \\
\bottomrule
\end{tabular}
\end{document}